\pdfoutput=1
\documentclass[11pt,twoside,a4paper,cmspaper,final,collab]{cms-tdr}
\def\svnVersion{dae8952}\def\svnDate{2024/02/09}\def\cmsCernNoTag{CERN-EP-2023-164}\def\cmsCernDate{\today}\def\cmsMessage{Published in Physics Letters B as \href{http://dx.doi.org/10.1016/j.physletb.2024.138518}{\doi{10.1016/j.physletb.2024.138518}.}}
\begin{document}\cmsNoteHeader{HIN-21-001}

\newcommand{\mumu}{\ensuremath{\PGmp\PGmm}\xspace}
\newcommand{\mmumu}{\ensuremath{m_{\mumu}}\xspace}
\newcommand{\NTrg}{\ensuremath{N_{\text{trig}}}\xspace}
\newcommand{\NBkgd}{\ensuremath{N_{\text{Bkg}}}\xspace}
\newcommand{\NAsso}{\ensuremath{N_{\text{assoc}}}\xspace}
\newcommand{\NPair}{\ensuremath{N_{\text{pair}}}\xspace}
\newcommand{\NSigone}{\ensuremath{N_{1\mathrm{S}}}\xspace}
\newcommand{\NSigtwo}{\ensuremath{N_{2\mathrm{S}}}\xspace}
\newcommand{\NSigthree}{\ensuremath{N_{3\mathrm{S}}}\xspace}
\newcommand{\ntrkoff}{\ensuremath{N^{\text{offline}}_{\text{trk}}}\xspace}
\newcommand{\sqrts}{\ensuremath{\sqrt{s}}\xspace}
\newcommand{\vtwo}{\ensuremath{v_{2}}}
\newcommand{\vn}{\ensuremath{v_{n}}}
\newcommand{\Vtwo}{\ensuremath{V_{2}}}
\newcommand{\vsub}{\ensuremath{v_{2}^{\text{sub}}}}
\newcommand{\Vsub}{\ensuremath{V_{2}^{\text{sub}}}}
\newcommand{\Jjet}{\ensuremath{J_{\text{jet}}}}
\newcommand{\pttrg}{\ensuremath{\pt^{\text{trig}}}}
\newcommand{\ptasso}{\ensuremath{\pt^{\text{assoc}}}}
\newcommand{\aeta}{\ensuremath{\abs{\eta}}}
\newcommand{\aetamu}{\ensuremath{\abs{\eta^{\PGm}}}}
\newcommand{\Deta}{\ensuremath{\Delta\eta}}
\newcommand{\Dphi}{\ensuremath{\Delta\phi}}
\newcommand{\rf}{\ensuremath{\text{ref}}}

\ifthenelse{\boolean{cms@external}}{\providecommand{\cmsLeft}{upper\xspace}}{\providecommand{\cmsLeft}{left\xspace}}
\ifthenelse{\boolean{cms@external}}{\providecommand{\cmsRight}{lower\xspace}}{\providecommand{\cmsRight}{right\xspace}}

\cmsNoteHeader{HIN-21-001}
\title{Study of azimuthal anisotropy of \texorpdfstring{\PgUa}{Upsilon(1S)} mesons in pPb collisions at \texorpdfstring{\sqrtsNN = 8.16\TeV}{sqrt(s[NN]) = 8.16 TeV}}

\date{\today}

\abstract{
   The azimuthal anisotropy of \PgUa mesons in high-multiplicity proton-lead collisions is studied using data collected by the CMS experiment at a nucleon-nucleon center-of-mass energy of 8.16\TeV. The \PgUa mesons are reconstructed using their dimuon decay channel. The anisotropy is characterized by the second Fourier harmonic coefficients, found using a two-particle correlation technique, in which the \PgUa mesons are correlated with charged hadrons. A large pseudorapidity gap is used to suppress short-range correlations. Nonflow contamination from the dijet background is removed using a low-multiplicity subtraction method, and the results are presented as a function of \PgUa transverse momentum. The azimuthal anisotropies are smaller than those found for charmonia in proton-lead collisions at the same collision energy, but are consistent with values found for \PgUa mesons in lead-lead interactions at a nucleon-nucleon center-of-mass energy of 5.02\TeV.
}

\hypersetup{
pdfauthor={CMS Collaboration},
pdftitle={Study of azimuthal anisotropy of Upsilon(1S) mesons in pPb collisions at sqrt(s[NN]) = 8.16 TeV},
pdfsubject={CMS},
pdfkeywords={CMS, pPb, heavy ion, v2, upsilon, flow}}

\maketitle 

\section{Introduction}
\label{sec:introduction}

Substantial azimuthal ($\phi$) correlations are observed among particles emitted over a wide pseudorapidity ($\eta$) range in heavy ion collisions at the BNL RHIC~\cite{star:v2_AuAu_130G, phenix:v2_AuAu_130G, phobos:v2_AuAu, star:high_energy_v2_n_jet, phobos:high_pt_v2} and the CERN LHC~\cite{cms:Long-range, cms:anisotropy, alice:hadron_v2_2p76, atlas:hadron_v2_2p76, cms:hadron_v2_2p76, cms:ultra_cent_hadron_v2_2p76}. These correlations are understood to arise from a nearly ideal hydrodynamic behavior of the strongly interacting quark-gluon plasma (QGP) created in such collisions~\cite{cite:coll_flow, cite:viscosity, cite:hydrodynamic_model}.
Similar long-range correlations are also observed in high particle multiplicity events in smaller colliding systems, such as 
proton-proton (pp)~\cite{cms:Obs-Long-range-pp, atlas:pp_v2_13Tev, cms:pp_v2_13Tev, cms:EvdCollectivity-pp},
proton-lead (pPb)~\cite{cms:Obs-Long-range-ppb, alice:pPb_v2_5TeV, atlas:pPb_v2_5TeV, lhcb:pPb_v2_5TeV, alice:pPb_pi_K_v2_5TeV, cms:pPb_PbPb_strange_v2, cms:pPb_multpart_v2, atlas:pp_pPb_PbPb_multpart_v2, atlas:pp_pPb_multpart_cumulant_v2},
proton-gold~\cite{phenix:pAu_200GeV_v2}, deuteron-gold~\cite{star:dAu_200GeV_v2}, and $^{3}$He--gold~\cite{phenix:HeAu_200GeV_v2, phenix:lightAu_200GeV_v2}. The origin of the observed correlations in light systems is still under active investigation~\cite{cms:multi_flow_pPb_8TeV, theory:smallsystem_femtoscopy, cms:c_b_flow_pp_pPb, theory:QGP_flow_overview}.

Heavy-flavor quarks (charm and bottom) are produced in the early stages of heavy ion collisions via hard scattering processes. 
Therefore, they experience the entire evolution of the produced medium~\cite{cite:cc_bb_in_NN}, making them useful probes to study the properties of the created matter.
The effect of the QGP medium on the propagating heavy-flavor quarks can be studied using the anisotropic azimuthal distribution of the emitted particles, which is experimentally characterized by its Fourier components. The second order coefficient, elliptic flow ($\vtwo$), reflects the anisotropy of the initial collision geometry~\cite{theory:flow_Fourier}.
Positive $\vtwo$ values have been reported for charmed mesons like $\PDz$ (open charm)~\cite{star:d0_v2_AuAu, phenix:d0_v2_AuAu,alice:D_v2_PbPb_2p76, alice:D_v2_PbPb_5p02, cms:D0_v2_PbPb_5p02} at RHIC and LHC, in gold-gold and lead-lead (PbPb) collisions, respectively, indicating the presence of collective flow for charm quarks. The situation for $\PJGy$ mesons (hidden charm) is less clear. In this case, the LHC results also support the presence of collective flow~\cite{cms:Jpsi_Raa_v2_PbPb,alice:Jpsiflow_PbPb}, but similar non-zero values were not observed at RHIC~\cite{star:Jpsi_v2_AuAu}.

Positive $\vtwo$ values have also been found in high-multiplicity pPb collisions for $\PDz$ and $\PJGy$ mesons by the CMS Collaboration~\cite{cms:Dflow-pPb_8TeV,cms:Jpsiflow-pPb}. 
Measurements by the ATLAS~\cite{atlas:mu_v2_pp_13TeV} and CMS Collaborations~\cite{cms:c_b_flow_pp_pPb} show indication of positive $\vtwo$ signals for charm quarks in high-multiplicity pp collisions. Also, muons for heavy-flavor hadron decays in high-multiplicity pPb collisions have positive $\vtwo$ values, as reported by the ALICE Collaboration~\cite{alice:mu_v2_pPb_8TeV}. These results suggest the possibility of collective flow for charm quarks in pp and pPb collisions.
Bottom quarks, on the other hand, show little, if any, sign of collective flow in either heavy or light collision systems. Both ALICE~\cite{alice:upsilon_v2_PbPb_5p02TeV} and CMS~\cite{cms:upsilon_v2_PbPb_5p02TeV} have found $\vtwo$ values for the bottom quark \PgUa meson in PbPb collisions which are consistent with zero. Also, a dissociation transport model calculation predicts zero $\vtwo$ for \PgUa in pPb collisions~\cite{theory:shinchon}. Measurements by CMS of $\PDz$ mesons arising from b hadron decays~\cite{cms:c_b_flow_pp_pPb} in pp and pPb collisions, and by ATLAS of the azimuthal anisotropy of muons from bottom hadron decays in pp collisions~\cite{atlas:mu_v2_pp_13TeV}, have also failed to find significant evidence of bottom quark flow. In order to further clarify the situation for heavy-quark flow in large and small systems, a measurement of the $\vtwo$ value for \PgUa mesons in pPb collisions will allow a more direct comparison of the flow of hidden charm and hidden bottom mesons in pPb. These measurements are particularly important in light of the uncertainty still remaining regarding the origin of collective behavior in lighter systems.

This Letter reports measurements of $\vtwo$ values for \PgUa mesons based on long-range, two-particle correlations in 
high-multiplicity pPb collisions at a center-of-mass energy per nucleon pair of $\sqrtsNN=8.16\TeV$. The $\vtwo$ coefficient for the \PgUa meson is determined over a wide transverse momentum (\pt) range. To correct for the residual contribution from nonflow effects, such as back-to-back jet-like correlations, the $\vtwo$ results are presented after subtracting correlations obtained from low-multiplicity pPb events.
The results for this analysis are tabulated in the HEPData record~\cite{HEPData}.

\section{The CMS detector}
\label{sec:detector}

The central feature of the CMS apparatus is a superconducting solenoid of 6$\unit{m}$ internal diameter, providing a magnetic field of 3.8\unit{T}. Within the solenoid volume are a silicon pixel and strip tracker, a lead tungstate crystal electromagnetic calorimeter, as well as a brass and scintillator hadron calorimeter, each composed of a barrel and two endcap sections. The hadron forward (HF) calorimeters, located at $3<\aeta<5$, extend the $\eta$ coverage provided by the barrel and endcap detectors. The HF calorimeters are azimuthally subdivided into $20^{\circ}$ modular wedges and further segmented to form $0.175{\times}0.175$ radians ($\Deta{\times}\Dphi$) towers. Muons are detected in gas-ionization chambers embedded in the steel flux-return yoke outside the solenoid within the coverage of $\aeta<2.4$.
Events of interest are selected using a two-tiered trigger system. The first level (L1), composed of custom hardware processors, uses information from the calorimeters and muon detectors to select events at a rate of about 100\unit{kHz} within a fixed latency of about 4\mus~\cite{CMS:2020cmk}. The second level, known as the high-level trigger (HLT), consists of a farm of processors running a version of the full event reconstruction software optimized for fast processing. The HLT reduces the event rate to about 1\unit{kHz} before data storage~\cite{CMS:2016ngn}.
A more detailed description of the CMS detector, together with a definition of the coordinate system used and the relevant kinematic variables, can be found in Ref.~\cite{cms:CMSdetector}.

\section{Event selection}
\label{sec:eventselection}

This analysis uses pPb collision data collected with the CMS detector in 2016, corresponding to an integrated luminosity of 186\nbinv~\cite{CMS-PAS-LUM-17-002, cms:Lumi}.
Events are selected by the Level-1 trigger requiring two muon candidates reconstructed in the muon detectors using the loosest possible selection criteria to maximize detection efficiency. Additional event selection criteria are applied to remove beam-gas interaction as discussed in~\cite{cms:dNdeta}. Also, at least one HF tower on each side is required to have transverse energy $\ET>3\GeV$ in order to reject non-hadronic collisions. Pileup events, where multiple overlapping collisions occur in a single bunch crossing, are rejected by requiring that only one primary vertex be found within 15\unit{cm} from the nominal interaction point along the beam axis and within 0.2\unit{cm} of the beam axis in the transverse direction.

The azimuthal anisotropy results are reported for high-multiplicity events in the range $70\leq\ntrkoff<300$, where $\ntrkoff$ is the number of reconstructed primary charged-particle tracks~\cite{cms:offline_track} with $\aeta<2.4$ and $\pt>0.4\GeVc$. This range corresponds to the 25\% of minimum bias events with the highest particle multiplicities. Events with $\ntrkoff<50$ (corresponding to the 40\% of events with the lowest particle multiplicities) are also used to estimate the possible contribution of residual back-to-back jet-like correlations.
Similar selection criteria and background corrections have been used in previous pp and pPb analyses~\cite{cms:flow_pp_pPb_8TeV, cms:multi_flow_pPb_8TeV, cms:two_four_corr-pPb-PbPb}. 

To keep only high-quality muons, a set of offline selection criteria are applied to select ``soft" muons~\cite{cms:tnp,cms:mu_perform}.
A soft muon is required to have a matched reconstructed track in the inner tracking system. This track must have at least six silicon tracker (pixel+strip) hits, with at least one of these hits occurring in the silicon pixel tracker. In addition, the distance of the closest approach of a muon track to the primary vertex must be less than 0.3$\unit{cm}$ and 20$\unit{cm}$ in the transverse and longitudinal directions, respectively. 
Also, the pairs of oppositely charged muons are fitted with a common vertex constraint, and the $\chi^{2}$ probability calculated by a Kalman vertex filter fit~\cite{theory:Kalman} is required to be greater than 1\%.
Lastly, the muons are selected in the kinematic range $\pt^{\mu}>3.5\GeVc$ and $\aetamu<2.4$ to ensure high efficiency for the \PgUa meson reconstruction. The $\aetamu<2.4$ range corresponds to the center-of-mass pseudorapidity range of $-2.87<\eta_{CM}^\mu<1.93$.

\section{Analysis}
\label{sec:analysis}

\subsection{Two-particle correlation technique}
\label{sec:correlation}

The \PGU signals are reconstructed using their dimuon decay channel.
The invariant mass of the selected dimuons is studied in the range 8--14\GeVcc, which covers the \PgUa, \PgUb, and \PgUc resonance peaks, and which is sufficient to ensure the stability of the description of the background dimuon distribution in the high-mass region~\cite{cms:Up_ratio}.

In calculating the invariant mass distribution and in determining the anisotropy of the detected particles, acceptance and efficiency corrections are applied event-by-event.
Acceptance is defined as the probability for both decay muons originating from an \PGU meson to have $\pt^{\PGm}>3.5\GeVc$ and $\aetamu<2.4$. The efficiency correction is determined as a function of \pt using the \PYTHIA 8.212~\cite{cms:intropythia8} Monte Carlo (MC) generator with tune CUETP8M1~\cite{cms:mtune} and its range is 0.74--0.98.
The \PYTHIA MC samples are embedded in pPb MC events generated by EPOS LHC~\cite{lhc:epos} that simulate underlying background events. The MC events are reweighted such that the \pt spectra of \PgUa mesons match the ones in pPb data. The efficiency values are found to not significantly depend on multiplicity, as also found in previous bottomonium and charmonium studies for pPb collisions~\cite{cms:Jpsi_pPb_5p02,cms:Up_pPb_5p02}.
The efficiency, calculated using a full simulation of the CMS detector based on \GEANTfour~\cite{cms:geant4}, gives the fraction of dimuons within the acceptance that are reconstructed, satisfy the trigger selection, and pass the analysis selection criteria described in Section~\ref{sec:eventselection}.
An additional correction to the dimuon efficiency is applied to take into account possible differences between data and MC simulation using the "tag-and-probe" method with muon pairs from $\PJGy$ meson decays~\cite{cms:tnp}.
This technique calculates the single-muon efficiency in both data and simulation by requiring a tight selection on one muon (the "tag" muon), which is then paired with the other muon that has looser selection criteria (the "probe" muon). The ratio of the single-muon efficiency in data to that in simulation is applied as an event-by-event weight to the dimuon efficiency in simulation.
The systematic uncertainty originating from the scale factor is calculated by varying its value by its uncertainty.

The $v_{n}$ harmonics for the \PGU particles are measured using the long-range ($\abs{\Deta}>1$) two-particle correlation technique previously discussed in Refs.~\cite{cms:Long-range,cms:anisotropy,cms:Obs-Long-range-ppb}.
The \PGU candidates (trigger particles) with transverse momentum $\pttrg$ in the ranges [0, 3], [3, 6], [6, 10], [10, 30]\GeVc are coupled with the charged tracks (associated particles) with transverse momentum $0.3<\ptasso<3\GeVc$ and pseudorapidity $\abs{\eta^{\text{assoc}}}<2.4$. The associated particle yields are corrected for the detector acceptance, reconstruction efficiency, and fractions of misidentified tracks~\cite{cms:EvdCollectivity-pp}.

The per-trigger-particle associated yield distribution is defined by
\begin{linenomath}
\begin{equation}\label{eq:corrfn}
    \frac{1}{\NTrg}\frac{\rd^{2}\NPair}{\rd\Deta\rd\Dphi} = B(0,0) \,
    \frac{S(\Deta,\Dphi)}{B(\Deta,\Dphi)}, 
\end{equation}
\end{linenomath}
where $\Deta$ and $\Dphi$ are differences in $\eta$ and $\phi$ of the pair, $\NTrg$ and $\NPair$ are the numbers of trigger particles and trigger-associated pairs, respectively, in the event.
This analysis procedure is repeated for each $\mmumu$ and \pt interval.
The S($\Deta$, $\Dphi$) distribution represents the per-trigger particle yield of trigger-associated pairs from the same event ($\mmumu$, \pt) range.
A mixed-event pair yield distribution, B($\Deta$, $\Dphi$), is constructed by pairing the \PGU candidates with charged-particle tracks from different events to account for the combinatorial background and pair-acceptance effects. The normalization factor, B(0, 0), is the mixed-events yield for both particles of the pair going in the same direction, corresponding to the maximum pair-acceptance.
The sample of mixed events is formed by randomly selecting ten events from the same multiplicity ($\ntrkoff$) and \pt ranges that fall within the same 2$\unit{cm}$ wide longitudinal primary vertex range as the \PGU candidate event.
The specific multiplicity range limits for the mixed events are [70, 80], [80, 90], [90, 110], [110, 150], [150, 300] and [0, 35], [35, 50] for high- and low-multiplicity events, respectively. The ranges are combined for the final reported results. These multiplicity ranges are selected to have similar \PGU candidate yields.
We also report a result for the extended \pt range of 0--30\GeVc, by combining the per-trigger-particle associated yield distribution from individual \pt intervals.

Both signal and background distributions are normalized by $\NTrg$ after summing up the pair density distributions and the total number of the trigger particles for all events.
To quantify the correlation structure, the two-dimensional (2D) distributions are projected onto 1D distributions in $\Dphi$ by averaging over the $\Deta$ range.

\subsection{Elliptic flow extraction}
\label{sec:v2extraction}

The dimuon invariant mass spectrum for the \PGU candidates is fitted using a signal function, Sig($\mmumu$), and a background function, Bkg($\mmumu$), to describe the \PGU signals and background dimuons, respectively. 
To take into account the different momentum resolutions of the muons in the endcap and barrel sections, a sum of two Crystal Ball (CB) functions~\cite{desy:crystalball} is used to model each \PGU state. The width parameters for the two CB functions are allowed to vary independently, while the mean and radiative-tail parameters are set to be equal.
All parameters are kept to be the same among the three \PGU states except for the mean and width parameters, which have values for the excited states, $\PgUb$ and \PgUc, that are equal to those for the \PgUa state multiplied by the PDG~\cite{Zyla:2020zbs} mass ratios to the \PgUa state. 
The Bkg($\mmumu$) is formed by taking the product of an exponential function, which models the combinatorial mass distribution in the high-mass region, and an error function, which models the saturation that occurs because of the muon \pt cutoff.
The yield parameters for the \PGU signals $\NSigone$, $\NSigtwo$, and $\NSigthree$, and the background dimuons $\NBkgd$ are left free in the fit. 
More details about the fitting procedure can be found in Ref.~\cite{cms:upsilon_v2_PbPb_5p02TeV}.

\begin{figure}[h]
\centering
	\includegraphics[width=0.48\textwidth]{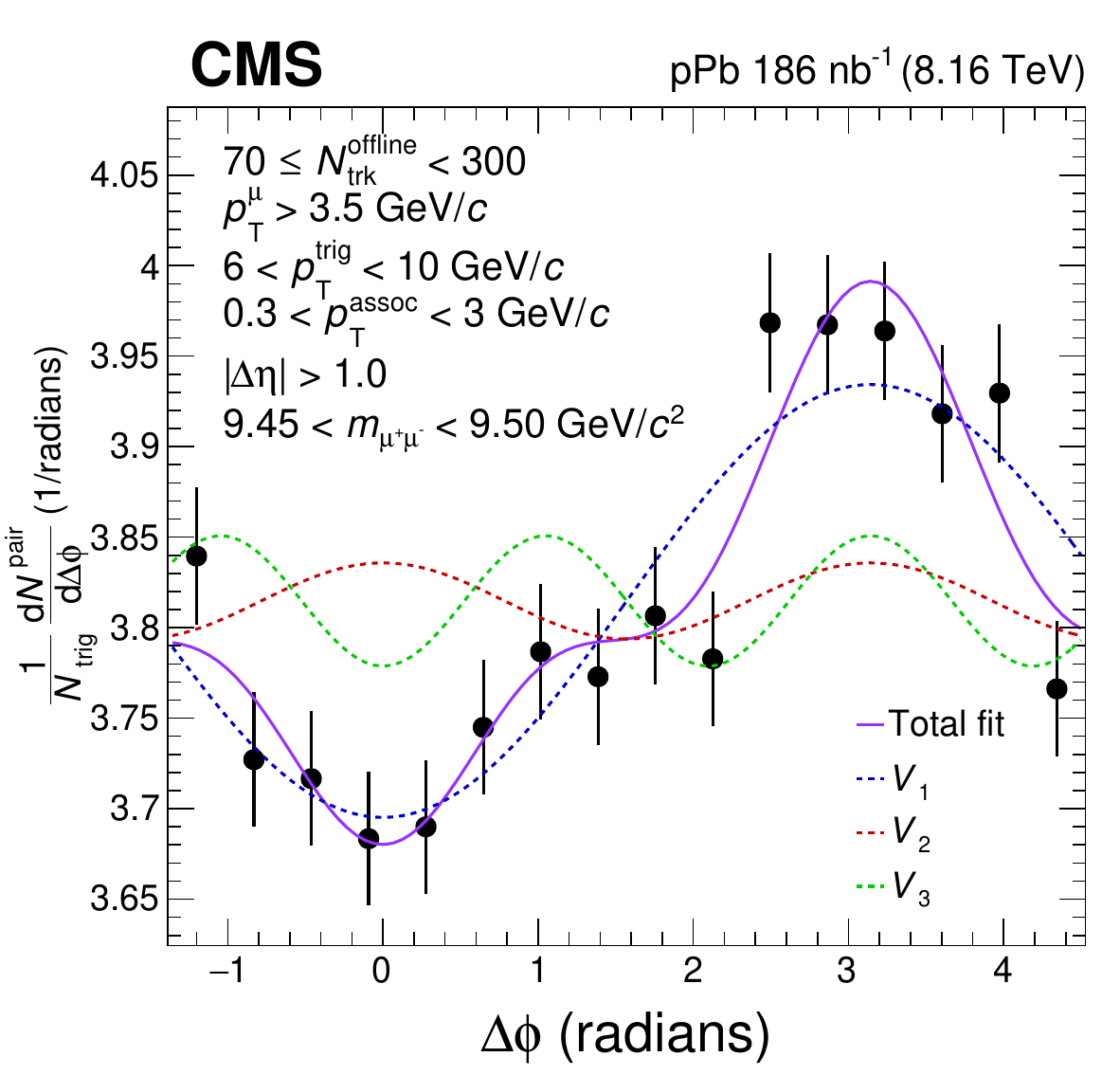}
	\caption{The $\Dphi$ projection of the \PgUa-track correlation function for the long-range correlation region ($\abs{\Deta}>1$) in high-multiplicity events ($70\leq\ntrkoff<300$). The three dashed lines represent the first three Fourier coefficients $V_{1}$ (blue), $\Vtwo$ (red), $V_{3}$ (green); the magenta solid line corresponds to the total fit.}
	\label{fig:fourierfitting}
\end{figure}

For the elliptic flow extraction, the azimuthal anisotropy harmonics are determined from a Fourier decomposition,
\begin{linenomath}
\begin{equation}\label{eq:corrfn_v2}
    \frac{1}{\NTrg}\frac{\rd\NPair}{\rd\Dphi} = 
    \frac{\NAsso}{2\pi} \left\{ 1+\sum_{n} 2V_{n} \cos(n\Dphi) \right\},
\end{equation}
\end{linenomath}
where $V_{n}$ are the Fourier coefficients and $\NAsso$ represents the total number of pairs per trigger particle for a given ($\pttrg,\ptasso$) range; 
$V_{n}$ is assumed to be the product of single-particle anisotropies of \PgUa mesons and reference charged particles~\cite{cms:two_four_corr-pPb-PbPb}.
The Fourier coefficients are studied across the dimuons invariant mass range, by dividing it into 23 intervals. The number of intervals is chosen to ensure similar yield in each $\mmumu$ interval. The number of intervals was varied and found to have little effect on the final results. The first three terms of the Fourier series $V_{n}^{\text{Sig+Bkg}}$, with n=1, 2, and 3, capture the most significant components and are determined for each $\mmumu$ interval using a binned $\chi^{2}$ fit. Figure~\ref{fig:fourierfitting} shows a typical $\Dphi$ projection fit.

The signal component of the $\Vtwo$ coefficient is found using a simultaneous binned $\chi^{2}$ fit of the invariant mass and $\Vtwo$ distribution, with
\begin{linenomath}
\label{eq:Simultaneousfunction}
\ifthenelse{\boolean{cms@external}}
{
\begin{multline}
  \Vtwo^{\text{Sig+Bkg}}(\mmumu) = \alpha_{1}(\mmumu)\Vtwo^{\PgUa} + \alpha_{2}(\mmumu)\Vtwo^{\PgUb}  \\
+ \alpha_{3}(\mmumu)\Vtwo^{\PgUc} + [ 1-\alpha_{1}(\mmumu)  \\
 - \alpha_{2}(\mmumu) - \alpha_{3}(\mmumu) ] v_{2}^{\text{Bkg}}(\mmumu),
\end{multline}
}
{
\begin{equation}
  \begin{aligned}
\Vtwo^{\text{Sig+Bkg}}(\mmumu) = \alpha_{1}(\mmumu)\Vtwo^{\PgUa} + \alpha_{2}(\mmumu)\Vtwo^{\PgUb} + \alpha_{3}(\mmumu)\Vtwo^{\PgUc} \\
    + [ 1-\alpha_{1}(\mmumu) - \alpha_{2}(\mmumu) - \alpha_{3}(\mmumu) ] \Vtwo^{\text{Bkg}}(\mmumu),
  \end{aligned}
\end{equation}
}
\end{linenomath}
where $\Vtwo^{\PgU(i\text{S})}$ ($i= 1, 2, 3$) is the $\Vtwo$ value for the $\PgU(i\text{S})$ mesons and is assumed to be independent of $\mmumu$.
The $\Vtwo$ distribution of the background dimuons as a function of mass, $\Vtwo^{\text{Bkg}}(\mmumu)$, is described by a second order polynomial function.
The $\alpha_{i}(\mmumu)$ factors are the signal fractions for the $\PgU(i\text{S})$ states as a function of $\mmumu$, as given by

\begin{figure}[ht]
\begin{centering}
	\includegraphics[width=0.48\textwidth]{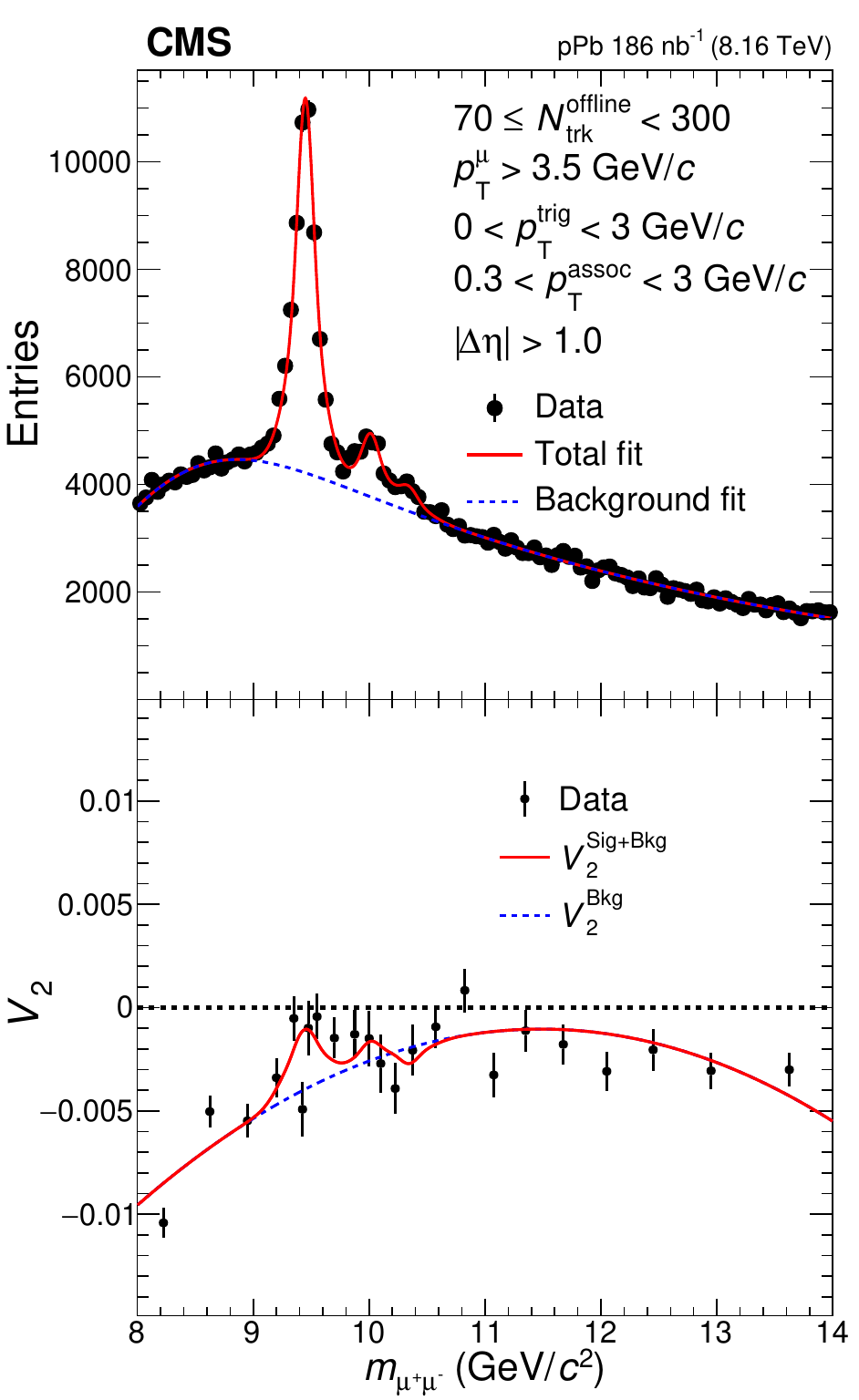}
	\caption{Simultaneous fit of the $\mmumu$ and $\Vtwo$ distributions for high-multiplicity ($70\leq\ntrkoff<300$) pPb events. The symbols indicate the event yields in the upper plot and $\Vtwo$ values from the Fourier component fit in the lower plot. The red solid (blue dash) lines correspond to the total (background) simultaneous fit to both distributions.}
	\label{fig:simulfit}
\end{centering}
\end{figure}

\begin{linenomath}
\label{eq:vnvsmasseq2}
\ifthenelse{\boolean{cms@external}}
{
\begin{multline}
\alpha_{i}(\mmumu) = \text{Sig}_{\PGUP{\textit{i}S}}(\mmumu)/[ \text{Sig}_{\PgUa}(\mmumu) \\
+ \text{Sig}_{\PgUb}(\mmumu) + \text{Sig}_{\PgUc}(\mmumu)\\
 + \text{Bkg}(\mmumu) ].
\end{multline}
}
{
\begin{multline}
\alpha_{i}(\mmumu) = \text{Sig}_{\PGUP{\textit{i}S}}(\mmumu)/[ \text{Sig}_{\PgUa}(\mmumu) \\
+ \text{Sig}_{\PgUb}(\mmumu) + \text{Sig}_{\PgUc}(\mmumu) + \text{Bkg}(\mmumu) ].
\end{multline}
}
\end{linenomath}
Figure~\ref{fig:simulfit} shows the fit to the mass spectrum and $\Vtwo^{\textrm{Sig+Bkg}}$ as a function of the invariant mass for dimuon pairs with $0<\pttrg<3\GeVc$ (the lowest \pt range) and event multiplicities $70\leq\ntrkoff<300$.

\subsection{Dijet subtraction}
\label{sec:subtraction}

The back-to-back nature of dijets can result in a significant $\Vtwo$ signal even if there is a large $\Deta$ gap between the \PgUa trigger particles and the associated charged hadrons. This nonflow contribution is subtracted using the low-multiplicity subtraction method developed in Refs.~\cite{cms:EvdCollectivity-pp, cms:two_four_corr-pPb-PbPb}.
The $\Vtwo^{\text{Sig}}$ values for $\ntrkoff<50$ are subtracted from the corresponding values in the higher multiplicity region, with
\begin{linenomath}
\ifthenelse{\boolean{cms@external}}
{
\begin{multline}
\label{eq:v2lowmultsub}
  \Vsub = \Vtwo^{\textrm{Sig}}(70\leq\ntrkoff<300) \\
 -\Vtwo^{\textrm{Sig}}(\ntrkoff<50) \\
 \times \frac{\NAsso(\ntrkoff<50)}{\NAsso(70\leq\ntrkoff<300)} \\
 \times \frac{\Jjet(70\leq\ntrkoff<300)}{\Jjet(\ntrkoff<50)},
\end{multline}
}
{
\begin{multline} 
\label{eq:v2lowmultsub}
  \Vsub = \Vtwo^{\text{Sig}}(70\leq\ntrkoff<300)-\Vtwo^{\text{Sig}}(\ntrkoff<50)\\
 \times \frac{\NAsso(\ntrkoff<50)}{\NAsso(70\leq\ntrkoff<300)} \, \frac{\Jjet(70\leq\ntrkoff<300)}{\Jjet(\ntrkoff<50)},
\end{multline}
}
\end{linenomath}
where $\Jjet$ represents the near-side jet yield, which is defined as the integral in the short-range region $(\abs{\Deta}<1)$ of the 1D $\Dphi$ correlation functions.
The associated ratio, $\NAsso(\ntrkoff<50)/\NAsso(70\leq\ntrkoff<300)$, is introduced to account for the enhanced jet yield due to the difference in the associated track yield regardless of the $\Deta$ gap.
The ratio, $\Jjet(70\leq\ntrkoff<300)/\Jjet(\ntrkoff<50)$, is introduced to account for the enhanced jet correlations resulting from the selection of higher-multiplicity events. 
For $\pt<10\GeVc$, where the \PgUa candidates tend to have values of $\eta$ outside the acceptance for charged particles (of $\aeta<2.4$), the jet yield ratio cannot be directly estimated from the two-particle azimuthal correlations.
In this case, based on previous analyses in the charm sector where no \pt dependence was observed~\cite{cms:Jpsiflow-pPb, cms:Dflow-pPb_8TeV}, the jet yield ratio value is assumed to be the same as that for the high-\pt region.
The \PgUa $\Vtwo$ results, after subtracting the correlations obtained from low-multiplicity events, are denoted as $\Vsub$.

To determine the \PgUa $\vtwo$ values, factorization is assumed, where the two-particle $\Vtwo$ value is taken as the product of single particle $\vtwo$ values for the trigger \PgUa particle and the associated charged hadrons, with
\begin{linenomath}
\begin{equation}\label{eq:corrfn_v2_trksub}
	\vsub(\pttrg) = \frac{\Vsub(\pttrg,\ptasso)}{\sqrt{\Vsub(\ptasso,\ptasso)}}.
\end{equation}
\end{linenomath}
The quantity $\Vsub(\ptasso,\ptasso)$ represents the Fourier coefficients extracted by correlating two reference charged particles.

\section{Systematic uncertainties}
\label{sec:systematics}

The systematic uncertainties from acceptance and efficiency corrections are estimated by comparing the $\vsub$ values with and without applying the corrections.
To account for the differences between MC simulation and data for the single-muon efficiency, an additional correction factor from the tag-and-probe technique is used.
The difference between the resulting $\vsub$ and the nominal $\vsub$ value is taken as the systematic uncertainty for this source.

To estimate the systematic uncertainty for the fit procedure, alternative functions are used instead of the nominal functions. A sum of Crystal Ball and Gaussian functions is used as an alternative function for the signal mass distribution.
For the background template, a fourth-order polynomial is used to replace the product of the error function and the exponential function. The background function in the $V^{\text{Sig+Bkg}}_{2}$ shape is also changed, from a second-order to a third-order polynomial.

In the simultaneous fitting step described in Section~\ref{sec:v2extraction}, all parameters for the Sig($\mmumu$) and Bkg($\mmumu$), except the yields for the \PGU states and the background dimuons, are fixed to achieve fit stability. The uncertainties associated with holding these fit parameters fixed are studied by releasing them, one by one, and redoing the fit.
The differences between the nominal $\vsub$ values and those where a parameter is allowed to vary are determined, and the root-mean-square average value is taken as the fit uncertainty. 

The final $\vsub$ value can be affected by the amount of low-multiplicity subtraction.
The systematic uncertainty associated with this effect is estimated by varying the jet ratio in Eq.~(\ref{eq:v2lowmultsub}) by $\pm3\sigma$ for $\pt<10\GeVc$, and by $\pm1\sigma$ for $\pt>10\GeVc$;
here, $\sigma$ is the standard deviation of the jet ratio, as obtained by fitting the $\Dphi$ distribution for $\Deta<1$.
In this analysis, jets are mostly produced in the same direction as the \PgUa particles. As the geometrical acceptance in the CMS detector for charged particles is $\aeta<2.4$, we lose part of the jet tracks (the ones at $\abs{\eta^{\textrm{trk}}}>2.4$) when the \PgUa is in the high-$\aeta$ region around 2.4.
This leads to a truncation of jets in the $\Deta\Dphi$ 2D distribution. Because such truncations are mostly found at low \pt, the jet yield ratio is calculated from the fit using only the high-\pt region. Hence, a larger variation is considered for the \pt ranges below 10\GeVc.

The response of the CMS detector to the MC generated particles was simulated using the \GEANTfour toolkit. To check the robustness of the full simulation, a closure test is performed by comparing generator- and reconstruction-level results in the MC simulation. The uncertainty associated with this closure test is in the range 0.002--0.007.

Individual systematic uncertainties are expected to be uncorrelated, and are therefore added in quadrature to obtain the final quoted values.
Systematic uncertainties from HLT trigger biases and pileup effects are assumed to be the same for \PgUa particles and charged particles and, consequently, they cancel in applying Eq.~(\ref{eq:corrfn_v2_trksub}).

\begin{table*}[htp]
	\centering
  \topcaption{Absolute systematic uncertainties ($10^{-2}$) included in the measurement of $\vsub$ evaluated in different $\pt^{\PgUa}$ ranges (in \GeVc).}
	\label{tab:systematics_list}
		\begin{tabular}{lcccc|c}
			\hline
			\multirow{2}{*}{Systematic source} & \multicolumn{5}{c}{$\pt^{\PgUa}$ range in \GeVc.} \\
			& [0-3] & [3-6] & [6-10] & [10-30] & [0-30] \\
			\hline
			Acceptance                  & 0.04 & 0.05 & 0.10 & 0.26 & 0.07 \\
			Efficiency                  & 0.01 & 0.01 & 0.02 & 0.05 & 0.05 \\
			Tag-and-probe technique     & 0.11 & 0.10 & 0.03 & 0.07 & 0.03 \\
			Signal function             & 0.06 & 0.03 & 0.02 & 0.21 & 0.10 \\
			Background function         & 0.03 & 0.01 & 0.13 & 0.59 & 0.12 \\
			$\Vtwo$ background function & 0.31 & 0.07 & 0.12 & 0.03 & 0.07 \\
			Parameter choice            & 0.06 & 0.01 & 0.04 & 0.07 & 0.01 \\
			Jet ratio                   & 0.32 & 0.43 & 2.18 & 1.88 & 0.45  \\
			MC closure test             & 0.66 & 0.23 & 0.35 & 0.32 & 0.50 \\
			\hline
			Total                       & 0.81 & 0.51 & 2.22 & 2.03 & 0.70 \\
			\hline
		\end{tabular}
	\centering
\end{table*}

The systematic uncertainties for each source are listed in Table~\ref{tab:systematics_list}. The total systematic uncertainty for the measured $\vsub$ values, taken as the sum in quadrature of all sources considered, is in the range 0.005--0.022, depending on the \PgUa meson \pt. Because of the low yield of dimuons in the highest \pt interval, the uncertainties for the signal extraction in the mass fit (signal function and background function) are found to be the largest for this interval. The uncertainty is dominated by the MC closure test, the background function, the jet ratio, and the $\Vtwo$ background function.

\section{Results}
\label{sec:results}

\begin{figure*}[h]
\begin{centering}
	\includegraphics[width=0.45\textwidth]{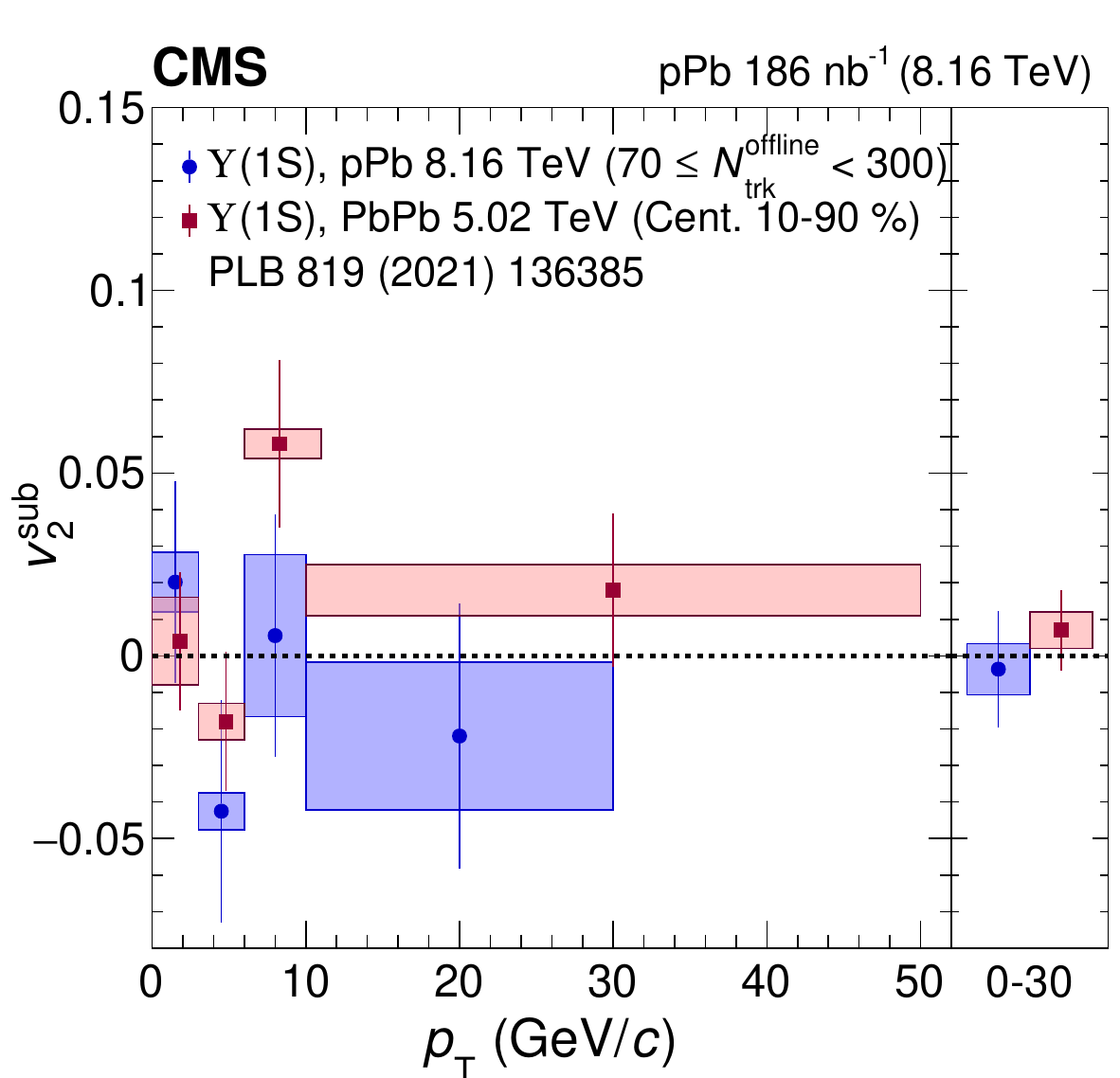}
	\includegraphics[width=0.45\textwidth]{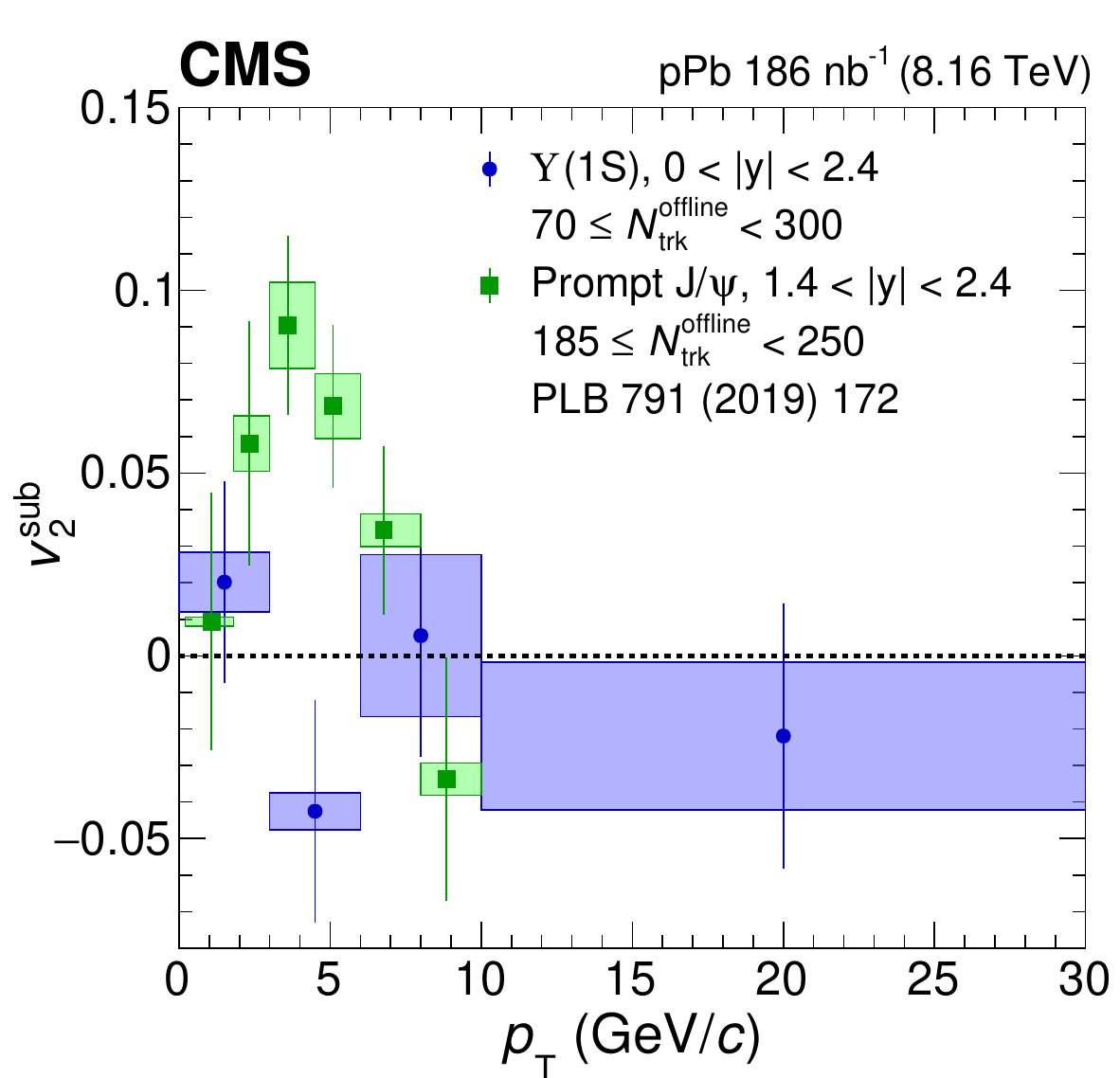}
	\caption{The \pt dependent $\vsub$ values of \PgUa mesons measured in the high-multiplicity region $70\leq\ntrkoff<300$, where a low-multiplicity region $\ntrkoff<50$ is used to estimate and correct for the dijet contribution. (Left) The results are compared to the corresponding results from PbPb collisions at 5.02\TeV, measured within the 10--90\% centrality range~\cite{cms:upsilon_v2_PbPb_5p02TeV}. The integrated range is 0--50\GeVc for PbPb. (Right) The same distribution is also compared with the $\vsub$ values for prompt $\PJGy$ mesons within $1.4<\abs{\textrm{y}}<2.4$ in pPb collisions at 8.16\TeV for $185\leq\ntrkoff<250$, where a low-multiplicity range with $\ntrkoff<35$ was used to estimate and correct for the dijet contribution~\cite{cms:Jpsiflow-pPb}. The vertical bars denote statistical uncertainties and the shaded boxes represent systematic uncertainties, while the widths of the boxes represent the \pt ranges.}
	\label{fig:compv2}
\end{centering}
\end{figure*}

Figure~\ref{fig:compv2} shows the comparison of the current $\vsub$ results for \PgUa mesons in pPb collisions with those for \PgUa in PbPb (left panel)~\cite{cms:upsilon_v2_PbPb_5p02TeV} and $\PJGy$ in pPb (right panel)~\cite{cms:Jpsiflow-pPb}, as a function of the \PgUa meson \pt.
The symbols are plotted at the center of each \pt range. The average \pt for the highest range is found to be around 15\GeVc. The observed $\vsub$ values for \PgUa in pPb are consistent with zero over the measured \pt range, as also found for centrality-integrated PbPb data.
This suggests that the azimuthal anisotropy of \PgUa mesons does not significantly depend on the size of the system despite the different level of suppression in pPb and PbPb collisions~\cite{cms:Up_pPb_5p02}. Future $\vtwo$ measurements for the excited states with larger data sets would be an interesting subject to further study the medium response of bottomonia in small collision systems.
In Fig.~\ref{fig:compv2} (right), we also compare our results with those of $\PJGy$ mesons, which exhibit significant flow-related correlations for the range of $0<\pt<10\GeVc$~\cite{cms:Jpsiflow-pPb}.
It should be noted that the $\ntrkoff$ ranges are different for the $\PJGy$ and \PgUa measurements. Restricting the \PgUa study to an $\ntrkoff$ range similar to that for the $\PJGy$ analysis does not significantly change the observed \PgUa distribution, although it does increase the statistical uncertainties.
In contrast to the \PgUa results, positive $\vsub$ values are reported for $\PJGy$ mesons in the range $2<\pt<8\GeVc$. To investigate the variation in $\vsub$ for these two quarkonium species, a p-value comparison is employed. The p-value determined for each $\PJGy$ $\vsub$ value against the hypothesis of an equivalent $\vsub$ value for \PgUa mesons in the \pt integrated range, taking into account the given uncertainties. Notably, in the range $2<\pt<8\GeVc$, p-values less than 0.05, indicating dissimilar results, are observed for the two $\PJGy$ \pt intervals within the range $3<\pt<6\GeVc$, with p-values of 0.0035 and 0.0150, respectively.
This indicates that bottom quarks experience less collective motion, if any, than charm quarks in pPb collisions.

\section{Summary}
\label{sec:summary}

The azimuthal anisotropy of \PgUa mesons, as quantified using the elliptic flow coefficients $\vsub$, is measured as a function of transverse momentum for the first time in high-multiplicity proton-lead (pPb) collision events at the center-of-mass energy per nucleon pair of $\sqrtsNN=8.16\TeV$. The measurement is corrected for dijet background using a low-multiplicity subtraction method. 
The $\vsub$ values are observed to be consistent with zero.
These results are also consistent with those found previously in PbPb collisions at $\sqrtsNN=5.02\TeV$.
Comparing $\vsub$ of \PgUa and those of $\PJGy$ mesons, the results suggest that any collectivity of bottomonia is smaller than that of charmonia in pPb collisions.
The direct comparison of \PgUa flow results for pPb and PbPb reinforces the conclusion that there is no evidence for collective flow of bottom quarks in systems of any size.

\begin{acknowledgments}
	We congratulate our colleagues in the CERN accelerator departments for the excellent performance of the LHC and thank the technical and administrative staffs at CERN and at other CMS institutes for their contributions to the success of the CMS effort. In addition, we gratefully acknowledge the computing centres and personnel of the Worldwide LHC Computing Grid and other centres for delivering so effectively the computing infrastructure essential to our analyses. Finally, we acknowledge the enduring support for the construction and operation of the LHC, the CMS detector, and the supporting computing infrastructure provided by the following funding agencies: SC (Armenia), BMBWF and FWF (Austria); FNRS and FWO (Belgium); CNPq, CAPES, FAPERJ, FAPERGS, and FAPESP (Brazil); MES and BNSF (Bulgaria); CERN; CAS, MoST, and NSFC (China); MINCIENCIAS (Colombia); MSES and CSF (Croatia); RIF (Cyprus); SENESCYT (Ecuador); MoER, ERC PUT and ERDF (Estonia); Academy of Finland, MEC, and HIP (Finland); CEA and CNRS/IN2P3 (France); SRNSF (Georgia); BMBF, DFG, and HGF (Germany); GSRI (Greece); NKFIH (Hungary); DAE and DST (India); IPM (Iran); SFI (Ireland); INFN (Italy); MSIP and NRF (Republic of Korea); MES (Latvia); LAS (Lithuania); MOE and UM (Malaysia); BUAP, CINVESTAV, CONACYT, LNS, SEP, and UASLP-FAI (Mexico); MOS (Montenegro); MBIE (New Zealand); PAEC (Pakistan); MES and NSC (Poland); FCT (Portugal); MESTD (Serbia); MCIN/AEI and PCTI (Spain); MOSTR (Sri Lanka); Swiss Funding Agencies (Switzerland); MST (Taipei); MHESI and NSTDA (Thailand); TUBITAK and TENMAK (Turkey); NASU (Ukraine); STFC (United Kingdom); DOE and NSF (USA).
	
	\hyphenation{Rachada-pisek} Individuals have received support from the Marie-Curie programme and the European Research Council and Horizon 2020 Grant, contract Nos.\ 675440, 724704, 752730, 758316, 765710, 824093, and COST Action CA16108 (European Union); the Leventis Foundation; the Alfred P.\ Sloan Foundation; the Alexander von Humboldt Foundation; the Science Committee, project no. 22rl-037 (Armenia); the Belgian Federal Science Policy Office; the Fonds pour la Formation \`a la Recherche dans l'Industrie et dans l'Agriculture (FRIA-Belgium); the Agentschap voor Innovatie door Wetenschap en Technologie (IWT-Belgium); the F.R.S.-FNRS and FWO (Belgium) under the ``Excellence of Science -- EOS" -- be.h project n.\ 30820817; the Beijing Municipal Science \& Technology Commission, No. Z191100007219010 and Fundamental Research Funds for the Central Universities (China); the Ministry of Education, Youth and Sports (MEYS) of the Czech Republic; the Shota Rustaveli National Science Foundation, grant FR-22-985 (Georgia); the Deutsche Forschungsgemeinschaft (DFG), under Germany's Excellence Strategy -- EXC 2121 ``Quantum Universe" -- 390833306, and under project number 400140256 - GRK2497; the Hellenic Foundation for Research and Innovation (HFRI), Project Number 2288 (Greece); the Hungarian Academy of Sciences, the New National Excellence Program - \'UNKP, the NKFIH research grants K 124845, K 124850, K 128713, K 128786, K 129058, K 131991, K 133046, K 138136, K 143460, K 143477, 2020-2.2.1-ED-2021-00181, and TKP2021-NKTA-64 (Hungary); the Council of Science and Industrial Research, India; ICSC -- National Research Centre for High Performance Computing, Big Data and Quantum Computing, funded by the EU NexGeneration program (Italy); the Latvian Council of Science; the Ministry of Education and Science, project no. 2022/WK/14, and the National Science Center, contracts Opus 2021/41/B/ST2/01369 and 2021/43/B/ST2/01552 (Poland); the Funda\c{c}\~ao para a Ci\^encia e a Tecnologia, grant CEECIND/01334/2018 (Portugal); the National Priorities Research Program by Qatar National Research Fund; MCIN/AEI/10.13039/501100011033, ERDF ``a way of making Europe", and the Programa Estatal de Fomento de la Investigaci{\'o}n Cient{\'i}fica y T{\'e}cnica de Excelencia Mar\'{\i}a de Maeztu, grant MDM-2017-0765 and Programa Severo Ochoa del Principado de Asturias (Spain); the Chulalongkorn Academic into Its 2nd Century Project Advancement Project, and the National Science, Research and Innovation Fund via the Program Management Unit for Human Resources \& Institutional Development, Research and Innovation, grant B05F650021 (Thailand); the Kavli Foundation; the Nvidia Corporation; the SuperMicro Corporation; the Welch Foundation, contract C-1845; and the Weston Havens Foundation (USA).
\end{acknowledgments}

\bibliography{auto_generated}
\cleardoublepage \appendix\section{The CMS Collaboration \label{app:collab}}\begin{sloppypar}\hyphenpenalty=5000\widowpenalty=500\clubpenalty=5000
\cmsinstitute{Yerevan Physics Institute, Yerevan, Armenia}
{\tolerance=6000
A.~Tumasyan\cmsorcid{0009-0000-0684-6742}
\par}
\cmsinstitute{Institut f\"{u}r Hochenergiephysik, Vienna, Austria}
{\tolerance=6000
W.~Adam\cmsorcid{0000-0001-9099-4341}, J.W.~Andrejkovic, T.~Bergauer\cmsorcid{0000-0002-5786-0293}, S.~Chatterjee\cmsorcid{0000-0003-2660-0349}, K.~Damanakis\cmsorcid{0000-0001-5389-2872}, M.~Dragicevic\cmsorcid{0000-0003-1967-6783}, A.~Escalante~Del~Valle\cmsorcid{0000-0002-9702-6359}, P.S.~Hussain\cmsorcid{0000-0002-4825-5278}, M.~Jeitler\cmsAuthorMark{1}\cmsorcid{0000-0002-5141-9560}, N.~Krammer\cmsorcid{0000-0002-0548-0985}, L.~Lechner\cmsorcid{0000-0002-3065-1141}, D.~Liko\cmsorcid{0000-0002-3380-473X}, I.~Mikulec\cmsorcid{0000-0003-0385-2746}, P.~Paulitsch, F.M.~Pitters, J.~Schieck\cmsAuthorMark{1}\cmsorcid{0000-0002-1058-8093}, R.~Sch\"{o}fbeck\cmsorcid{0000-0002-2332-8784}, D.~Schwarz\cmsorcid{0000-0002-3821-7331}, M.~Sonawane\cmsorcid{0000-0003-0510-7010}, S.~Templ\cmsorcid{0000-0003-3137-5692}, W.~Waltenberger\cmsorcid{0000-0002-6215-7228}, C.-E.~Wulz\cmsAuthorMark{1}\cmsorcid{0000-0001-9226-5812}
% ...
\par}
\vskip\cmsinstskip
\dag:~Deceased\\
$^{1}$Also at TU Wien, Vienna, Austria\\
% ...
\end{sloppypar}
\end{document}